\documentclass[11pt,authoryear]{article}
\usepackage[round]{natbib}
\usepackage{amsmath, amsthm, amssymb, amsfonts}
\usepackage{graphicx}
\usepackage{epstopdf}
\usepackage{epsfig,color}
\usepackage[utf8]{inputenc}
\usepackage[margin=3cm]{geometry}
\usepackage{color}
\usepackage{url}
\newcommand{\tr}{^{\prime}}
\def\bd#1{\mbox{\boldmath $#1$}}
\def\bdl#1{\mbox{\boldmath {\scriptsize $#1$}}}
\newcommand{\diag}{{\rm diag}}
\def\m#1{\mbox{#1}}                % - \m: testo in formula
\def\cg#1{\mbox{${\cal #1}$}}      % caratteri calligrafici

\newtheorem{lemma}{Lemma}

\newtheorem{proposition}{Proposition}
\newtheorem{example}{Example}
\newtheorem{remark}{Remark}
\author{Antonio Forcina, Department of Economics,\\
University of Perugia, Italy\\
Francesco Bartolucci, Department of Economics,\\
University of Perugia, Italy}
\title{Estimating the size of a closed population by modeling
latent and observed heterogeneity}
{\markboth{A. Forcina, F. Bartolucci}{Capture-recapture}

\begin{document}
\maketitle
\begin{abstract}
The paper describes a new class of capture-recapture models for closed populations when individual covariates are available. The novelty consists in combining a latent class model for the distribution of the capture history, where the class weights and the conditional distributions given the latent may depend on covariates, with a model for the marginal distribution of the available covariates as in \cite{Liu2017}. In addition, any general form of serial dependence is allowed when modeling capture histories conditionally on the latent and covariates.
A Fisher-scoring algorithm for maximum likelihood estimation is proposed, and the Implicit Function Theorem is used to show that the mapping between the marginal distribution of the observed covariates and the probabilities of being never captured is one-to-one. Asymptotic results are outlined, and a procedure for constructing likelihood based confidence intervals for the population size is presented. Two examples based on real data are used to illustrate the proposed approach
\paragraph{Keywords}
Capture-recapture data; Implicit function Theorem; Latent class models;  Likelihood based confidence intervals; Logistic models; Log-linear models; Recursive dependence.
\end{abstract}
\section{Introduction}
\label{sec:intro}
The literature on capture-recapture methods for estimating the size of populations not directly observable is wide; a recent overview is provided in \cite{BohningBook}, a coauthored book with contributions on the most important branches of the subject.
The present paper deals with situations where units belonging to a closed population
may be detected by a set of lists (or capture occasions) and the full capture history for each unit is available together with a set of individual covariates which may affect the probability of being detected by the different lists.

There is substantial agreement that likelihood based confidence intervals for the population total are preferable to intervals based on the normal approximation, when the sample size is not sufficiently large \citep[see, among others,][]{evans1994general}. The likelihood based approach for confidence interval has been dealt in several papers,
starting with   \cite{cormack1992interval}
whose method has been extended by \cite{StangPvdH} to the case of unobserved heterogeneity, modeled by latent classes, to construct intervals for the population  size within a specified sub-population indexed by a discrete covariate.
The approach proposed by \cite{Bart2006} also assumes latent class models and is designed to provide confidence intervals for the population size within a specified collection of sub-populations indexed, again, by a discrete covariate.
More recently, \cite{Liu2017} have proposed an approach which takes also into accounts the marginal distribution of the covariates; because a non-parametric model is used for the latter component, connections with \cite{Owen88}'s Empirical Likelihood are exploited.
% based on a semi-parametric model to account for the marginal distribution of covariates; since they assume that covariates are continuous, tools from \cite{Owen88}'s Empirical Likelihood ratio are used.

This paper extends \cite{Liu2017}'s work, which relies on models for the distribution of the capture history that only account for observed heterogeneity and assume independence between capture occasions, by considering unobserved heterogeneity in the capture probabilities through a latent class model with parameters that may depend on the covariates; in addition, flexible models accounting for the capture history conditional on the latent variable and observable covariates are presented. Among other features, our models allow pairs of lists to be dependent conditionally on the latent class as in a log-linear model, or to exhibit serial dependence, leading to models for which the conditional distribution of the capture configuration given the latent class is not log-linear in the parameters.
In addition, we provide an alternative to the Empirical Likelihood approach in \cite{Liu2017} to compute a profile likelihood for the parameters of interest. Our approach highlights the relation between the marginal distribution of the covariates and the conditional probability of being never captured which we show to be one-to-one.
%We also show that the marginal distribution of the covariates is an invertible function of the probability of being never captured, a result that provides an alternative to the Empirical Likelihood approach in \cite{Liu2017} to compute a profile likelihood for the parameters of interest by removing the marginal distribution of the covariates.

The remainder of the paper is structured as follows. In the next section we describe the proposed class of models. Maximum likelihood estimation is described in Section \ref{sec:mle} and asymptotic properties of the estimator are outlined in Section \ref{sec:asymp}. Finally, an illustration consisting of two applications based on data about human and animal populations is provided in Section \ref{sec:application}.
Software for fitting the models described in the paper, both in {\tt R} and {\sc Matlab}, is available from the authors.
\section{The proposed class of models}
\subsection{Preliminaries}
\label{}
Let $N$ denote the overall population size, $n$ be the number of subjects observed at least once, and suppose that, for each observed subject, a vector $\bd x$ of covariates is available. Let $s$ denote the number of distinct covariate configurations and $n_i$, $i=1,\dots, s$, be the number of subjects captured within the stratum having vector of covariates $\bd x=\bd x_i$, so that  $\sum_{i=1}^s n_i=n$. Though, in practice, covariates will always be discrete, if they can assume a large number of distinct configurations, we might have $n_i=1, \forall i$, meaning that each stratum contains just one subject; in general $s\leq n$.

%Let $\tau_i$ denote the point mass, in the sense of \cite{Owen88}, attached to observations with $\bd x=\bd x_i$ in
%the sample
Following \cite{Liu2017}, let $\tau_i$ denote the marginal weight attached to individuals with $\bd x=\bd x_i$ in the population
and $\phi_i$ be the probability that an individual with covariates $\bd x_i$ is never captured. The probability of being never captured for a subject whose covariate configuration is unknown may be computed as $\phi$ = $\sum_i \phi_i\tau_i$; because $\sum_i\tau_i=1$, this may be seen as a weighted average of the $\phi_i$s.

Assuming that population units may be captured by $J$ different lists, the capture history may be coded into a binary vector $\bd h$ = $(h_1, \dots , h_J)\tr$, where $h_j=1$ if a subject is captured by the $j$th list and 0 otherwise. We arrange the capture configurations in lexicographic order starting with $\bd h=\bd 0$, where $\bd 0$ denotes a column vector of 0s of suitable dimension. Let $k$ = $2^J$ denote the number of possible configurations.

Suppose that subjects can belong to $C$ disjoint latent classes and let $\pi_{ic,\bdl h}$ denote the probability that subjects within the $i$th stratum belong to latent class $c$ and experience capture configuration $\bd h$ conditionally on $\bd x$ = $\bd x_i$; let $p_{i,\bdl h}$ denote the probability of capture configuration $\bd h$ conditionally on the covariates and
$\bd y_i$ the vector whose entries are the frequencies for each capture configuration with the one corresponding to $\bd h=\bd 0$ excluded.
When $n_i$ = 1, namely only one subject with $\bd x=\bd x_i$ has been captured, $\bd y_i$ is a vector of 0s except for the entry corresponding to the observed capture configuration which equals 1.
Let also $\bd p_i$ and $\bd \pi_i$ denote the corresponding vectors of probabilities for the manifest and latent distributions respectively,  with the latent running the slowest and capture configurations in lexicographic order with the $\bd h=\bd 0$ configuration omitted.
Thus, $p_{i,\bdl 0}$ = $\phi_i$ = $1 - \bd 1\tr \bd p_i$, where, in general, by $\bd 1$ we denote a column vector of ones of suitable dimension.
\subsection{A latent class model for capture probabilities \label{FSLL}}
The set of models for capture data proposed here is a natural extension of the one described in \cite{Forc17};
relative to the class of models in \cite{Bart2006}, neither is a subset of the other, but the one proposed here is easier to estimate.
For a related class of models see also \cite{bartolucci2001analysis}.

Let $\bd\xi_i$ denote a $C$-dimensional column
vector whose entries are the marginal probabilities of the latent for the $i$th stratum and $\bd q_{ic}$ be a vector of multinomial probabilities that determine the conditional distribution of the responses conditionally on the latent equal to $c$, with the configuration $\bd h=\bd 0$ removed; let also $\bd Q_i$ be the matrix with columns $\bd q_{i1},\dots ,\bd q_{iC}$, so that $\bd p_i$ = $\bd Q_i\bd\xi_i$ corresponds to the manifest distribution of captures for the $i$th stratum.
Below we use the convention that a vector of multinomial probabilities including the response configuration $\bd h=\bd 0$ will be tilted.

As in \cite{Forc17}, we assume that the vector of latent weights follows a multinomial logistic regression model to account for the possible dependence of these weights on covariates; the model may be written as
$$
\log \bd\xi_i = \bd X_i\bd\zeta-\bd 1\log[\bd 1\tr\exp(\bd X_i\bd\zeta)],
$$
where $\bd X_i$ are suitable design matrices and $\bd\zeta$ is the corresponding vector of regression parameters.

Let $\tilde{\bd q}_{ic}$ denote the vector whose elements are the $q_{ic,\bdl h}$, having arranged the capture histories $\bd h$ in lexicographic order; we assume that %%%%%%%-------------%%%%%%%%%
$\tilde{\bd q}_{ic}$ = $\exp[\bd m_{ic}(\bd\delta_{ic})]$,
where $\bd m_{ic}(\bd\delta_{ic})$  is a vector of smooth functions of $\bd\delta_{ic}$, a working vector of parameters which, as we shall see, may depend on covariates and be subjected to linear restrictions. Log-linear models correspond to the special case
$$
\bd m_{ic}(\bd\delta_{ic}) = \bd G\bd\delta_{ic}-\log[\bd 1\tr\exp(\bd G\bd\delta_{ic})],
$$
where $\bd G$ is the design matrix of the log-linear model coding main effects (capture-occasions) and, possibly, a limited number of bivariate interactions.

To model recursive dependencies in a capture-recapture contest, let $\bd h_j$ denote the vector containing the first $j-1$ elements of $\bd h$, that is, the capture history up to occasion $j-1$, with $\bd h_1$ = $\emptyset$; let also $\eta_{icj,\bdl h_j}$ denote the logit of the probability of capturing the $i$th subject at the $j$th occasion, conditionally on the latent class and the previous capture history, and $q_{ic,\bdl h}$ denote the conditional probability of capture configuration $\bd h$ within latent class $c$.
The saturated model may be written as the product of the density of $J$ binary variables
$$
\log q_{ic,\bdl h} = \sum_{j=1}^J  \log q_{icj,\bdl h_j} =
\sum_{j=1}^J \left\{h_j \eta_{icj,\bdl h_j}- \log [1+\exp(\eta_{icj,\bdl h_j})]\right\},
$$
where $h_j$ is the $j$th element of $\bd h$; this formulation generalizes the logit model by \cite{Hggins1989} to latent heterogeneity and the models by \cite{farcomeni2011recapture} and \cite{FegaTa2016} to the case where observed and latent heterogeneity is present.
%the logit model by \cite{Hggins1989} accounts for dependence on past history and covariates, .
For fixed $i$ and $c$, the saturated model involves $k-1$ parameters; parsimonious models may be defined in two stages. First, like in \cite{FegaTa2016}, partition the set of all possible partial capture histories into $V$ disjoint sub-sets $(\cg H_1,\dots ,\cg H_V)$, where, usually, $\cg H_1$ corresponds to absence of capture history or no previous capture and assume that $\eta_{icj,\bdl h_j}$ = $\delta_{ic,v}$ for all $\bd h_j\in \cg H_v$.
This may be translated into a matrix formulation as follows: first, arrange the $k$ possible capture histories $\bd h\tr$ in lexicographic order into the rows of the matrix $\bd H$ and
%let $\bd H$ be the $k\times J$ binary matrix which has a row for each possible capture configuration, with these configurations arranged in lexicographic order;
let $\bd H_v$, $v=1,\dots , V$, be a matrix of the same size as $\bd H$ and such that the $j$th element in its $r$th row is equal to 1 if $\bd h_{j}$, the vector of partial history in the $r$th row of $\bd H$, belongs to $\cg H_v$ and 0 otherwise.
Because the equivalence classes are disjoint, if element $(r,j)$ of $\bd H_u$ is equal to 1, the corresponding elements of $\bd H_v$, $\forall v \neq u$, must be 0.
The non-zero elements of $\bd H_v$ might be replaced by quantitative summaries
%apart from 1, the elements of $\bd H_v$ may correspond to quantitative summaries
of the previous capture history as in \cite{FegaTa2016}.
%Let $\tilde{\bd q}_{ic}$ denote the vector whose elements are the $q_{ic,\bdl h}$, having arranged the capture histories $\bd h$ in lexicographic order;
Direct calculations show that we may write
$$
\log \tilde{\bd q}_{ic} = \bd A \bd\delta_{ic} - \bd B \log[1+\exp(\bd\delta_{ic})],
$$
where the matrices $\bd A$ and $\bd B$ may be constructed as follows: let '$*$' denote the element-wise product between matrices; then the $v$th column of $\bd A$ is $(\bd H*\bd H_v)\bd 1$ while the corresponding column of $\bd B$ is $\bd H_v\bd 1$.
Once a parsimonious model has been defined, linearity within the components of a recursive model, linear dependence on covariates and restrictions across latent classes may be defined as
$$
\bd\delta_{ic} = \bd M_{ic}\bd\lambda,\quad i=1,\ldots,s,\:c=1,\ldots,C.
$$
\begin{example}
A model slightly more complex than $M_b$ in \cite{otis1978statistical} which, for fixed $i$ and $c$, requires four parameters, may be defined by crossing the following two conditions, each of which may be true independently of the other: (i) captured more than once before and (ii) captured in the previous occasion. Thus we may have four types of capture histories: (1) at most once before, not in the previous occasion; (2) first time in the previous occasion; (3) more than once before, not in the previous occasion; (4) captured in the previous occasion and in at least one occasion before. An additive effect of these two features might be assumed.
\end{example}
\section{Maximum likelihood estimation}\label{sec:mle}
\subsection{The likelihood function}
By adapting the formulation in \cite{Liu2017} to the case of subjects possibly grouped into strata, the log-likelihood may be split into three components:
\begin{enumerate}
\item  the logarithm of the binomial likelihood, corresponding to the probability that $n$ out of $N$ subjects whose stratum is unknown are captured at least once, that is,
    $$
    \log \Gamma(N+1)-\log \Gamma(N-n+1)+(N-n)\log(\phi)+n\log(1-\phi);
    $$
\item the sum of the logarithm of the probability of the observed capture configurations conditionally on belonging to the $i$the stratum, $i=1,\dots ,s$, and on being captured at least once, equal to
    \begin{equation}
    \sum_{i=1}^s \bd y_i\tr\log\left(\frac{1}{1-\phi_i}\bd p_i\right);
    \label{Lcon}
    \end{equation}
\item the sum of the logarithm of the probability of observing $\bd x=\bd x_i$ conditionally on being captured at least once irrespective of the covariate value, times $n_i$, $i=1,\dots , s$, that is,
    $$
    \sum_{i=1}^s n_i\log\left(\tau_i\frac{1-\phi_i}{1-\phi}\right) = \sum_{i=1}^s n_i\log\tau_i+
    \sum_{i=1}^s n_i\log(1-\phi_i) -n\log(1-\phi).
    $$
\end{enumerate}
After simplifying terms appearing twice with opposite signs, the log-likelihood may be written as
\begin{equation}
L(\bd\psi,\bd\tau) = \log\frac{\Gamma(N+1)}{\Gamma(N-n+1)}+(N-n)\log(\bd\tau\tr\bd\phi)
+\sum_{i=1}^s\left[\bd y_i\tr\log(\bd p_i)+n_i\log \tau_i\right],
\label{LkSt}
\end{equation}
where $\bd\psi$ = $(N, \bd\beta\tr)\tr$, $\bd\beta$ = $(\bd\zeta\tr, \bd\lambda\tr)\tr$ and $\bd\phi$ is the $s$ dimensional vector with elements $\phi_i$.
To phrase the problem into an Empirical Likelihood context, \cite{Liu2017} treat $\phi$ (the weighted average of the $\phi_i$) as an independent parameter, while we make explicit the dependence of $\phi$ on $\bd\phi$ which is determined by $\bd\beta$.
%This formulation makes explicit the dependence of $\phi$ on $\bd\tau$, while \cite{Liu2017} treat it as an additional parameter.
Though the two approaches are equivalent, ours may be more easily implemented into a latent class context; in addition, as we show in Section \ref{sec:Implicit}, for fixed $N$ and $\bd\beta$, the maximum likelihood estimate (MLE) of $\bd\tau$ is an invertible function of $\bd\phi$. This result is useful for studying the profile likelihood of $N$ and $\bd\beta$ by treating $\bd\tau$ as a vector of nuisance parameters. We also derive an element-wise relationship between the $\tau_i$ and the $\phi_i$ parameters which has an interesting interpretation and allows comparisons with the similar relationship which holds with the conditional MLE.

Rather than maximizing the likelihood simultaneously with respect to the full set of parameters, we suggest updating the three components ($N,\:\bd\beta,\:\bd\tau$) separately and to iterate until convergence:
\begin{enumerate}
\item for fixed $N$ and $\bd\tau$, maximize $L(\bd\psi,\bd\tau)$ with respect to $\bd\beta$ by a Fisher-scoring algorithm;
\item for fixed $N$ and $\bd\beta$, maximize $L(\bd\psi,\bd\tau)$ with respect to $\bd\tau$ under the constraint $\bd\tau\tr\bd 1=1$;
\item for fixed $\bd\beta$ and $\bd\tau$ maximize $L(\bd\psi,\bd\tau)$ with respect to $N$.
\end{enumerate}
These steps are illustrated in detail in the following.
\subsection{Updating $\bd\beta$}
For fixed $N$ and $\bd\tau$, $\bd\beta$ can be updated by a Fisher-scoring algorithm similar to the one described in \cite{Forc17}.
Let $\bd\Phi$ denote the matrix of derivatives of $\bd\phi$ with respect to $\bd\beta\tr$; the score vector with respect to $\bd\beta$ may be written as
$$
\bd s_{\bdl\beta} = \frac{N-n}{\phi}\bd\Phi\tr \bd\tau + \sum_{i=1}^s\bd D_i\tr \diag(\bd p_i)^{-1}\bd y_i,
$$
where $\bd D_i$ is the derivative of $\bd p_i$ with respect to $\bd\beta\tr$ and may be partitioned as
$$
\bd D_i = (\bd Q_i\bd\Omega(\bd\xi_i)\bd X_i,\bd D_{i1},\dots ,\bd D_{iC})
$$
and, for $c=1,\dots ,C$, we have
\[
\begin{array}{ll}
\m{\it Log-linear } & \bd D_{ic} = \xi_{ic}\bd\Omega(\bd q_{ic})\bd M_{ic}, \quad  \\
\m{\it Recursive logistic } & \bd D_{ic} = \xi_{ic}\diag(\bd q_{ic})\left[\bd A-\bd B\diag(1+\exp(\bd\delta_{ic}))^{-1}
\exp(\bd\delta_{ic}) \right]\bd M_{ic},
\end{array}
\]
where, given a vector of multinomial probabilities $\bd v$, $\bd\Omega(\bd v)$ = $\diag(\bd v)-\bd v\bd v\tr$.
Lemma \ref{FInf} in the Appendix implies that the derivative of $\bd\Phi$ and $\bd D_i$ with respect to the elements of $\bd\beta$ may be neglected, thus the expected information matrix is easy to compute and takes the form
$$
\tilde{\bd F} = N\left[\bd\Phi\tr\bd\tau\bd\tau\tr\bd\Phi/\phi+\sum_{i=1}^s \tau_i\bd D_i\tr\diag(\bd p_i)^{-1}\bd D_i\right].
$$
\subsection{Estimation of $\bd\tau$} \label{sec:Implicit}
An alternative to the Empirical Likelihood approach for estimating $\bd\tau$ as in \cite{Liu2017} may be obtained by maximizing the log-likelihood as a function of $\bd\tau$ by Lagrange multipliers to account for the constraint $\bd 1\tr\bd\tau$ = 1. Let $\bd n$ denote the vector with entries $n_i$, $i=1,\ldots,s$, and $\gamma$ be the Lagrange multiplier; ignoring constant terms with respect to $\bd\tau$, we have
$$
L(\bd\tau) = (N-n)\log(\bd\phi\tr\bd\tau)+\bd n\tr\log(\bd\tau)-\gamma(\bd 1\tr\bd\tau-1).
$$
By differentiating with respect to $\bd\tau$ and a few simple algebraic manipulations, see the Appendix for details, we obtain the following updating equation where $u$ is the step counter
\begin{equation}
\bd\tau^{(u+1)} = \frac{1}{N}\left[\bd n+\frac{N-n}{\phi^{(u)}}\diag(\bd\phi)\bd\tau^{(u)}\right].
\label{tau-it}
\end{equation}
In practice, this equation will be iterated until $\sum_i \mid \tau_i^{(u-1)} -\tau_i^{(u)}\mid$
is sufficiently close to 0; formally, at convergence, $\tau_i^{(u-1)}$ = $\tau_i^{(u)}$ = $\tau_i$, $i=1,\dots, s$. With this in mind, by simple algebraic manipulations we obtain
\begin{equation}
\label{Hyper}
\tau_i = \frac{n_i\phi}{N\phi-(N-n)\phi_i}, \quad i=1,\dots s;
\end{equation}
though the right hand side still depends on $\bd\tau$ through $\phi$, at convergence this is a constant with respect to $i$ and, under the assumption that the elements of $\bd\phi$ are strictly positive, Lemma \ref{diti} in the Appendix imply that $N\phi-(N-n)\phi_i>0$, thus the points $(\phi_i,\tau_i)$ lie on the branch of an hyperbola which is  strictly increasing and concave.

Once written in matrix notation, (\ref{Hyper}) may be interpreted as an implicit equation in $\bd\phi$ and $\bd\tau$; to get a more convenient expression, multiply both sides by $N\phi\bd I-(N-n)\diag(\bd\phi)$,
rearrange terms and, for the moment, consider  the case of just one subject captured in each stratum, that is $\bd n$ = $\bd 1$; then we have
\begin{equation}
N\bd\tau\phi-\bd 1\phi-(N-n)\diag(\bd\tau)\bd\phi = \bd 0.
\label{Inplicit}
\end{equation}
An equivalent expression may be obtained from equation (11) in the Supplementary Material in \cite{Liu2017}. In order to derive an expression for $\bd D_{\bdl\phi}$ = $\partial\bd\tau /\partial\bd\phi\tr$, differentiate (\ref{Inplicit}) with respect to $\bd\phi\tr$ and rearrange terms (see the Appendix for details); we obtain
\begin{align}
\label{eq:Dphi}
& [\bd I + \bd a\bd\phi\tr\diag(\bd d_0)^{-1} ] \diag(\bd d_0) \bd D_{\bdl\phi} =
[\bd I- \bd a\bd\tau\tr\diag(\bd d_1)^{-1}]\diag(\bd d_1),\\
& \m{ where } \bd d_0=N\phi\left(\bd 1-\frac{N-n}{N\phi}\bd\phi\right),\:
\bd a=N(\bd\tau-\bd 1/N), \: \bd d_1 = (N-n)\bd\tau.
\end{align}
Let $\cg T$ denote the collection of pairs $(\bd\phi,\:\bd\tau)$ for which $\bd D_{\bdl\phi}$ exists; the implicit function theorem implies that $\bd\tau$ is an invertible function of $\bd\phi$ within $\cg T$. Lemma \ref{diti} in the Appendix shows that the elements of both $\bd a$ and $\bd d_0$ are strictly positive as long as the elements of $\bd\tau$ and $\bd\phi$ are also strictly positive. Thus, the following result holds.
\begin{proposition}
Within $\cg T$, $\bd\tau$ is an invertible function of $\bd\phi$ with derivative matrix $\bd D_{\bdl\phi}$.
\label{IFT}
\end{proposition}
\noindent{\sc Proof.} The left hand side of (\ref{eq:Dphi}) is well defined because the elements of $\bd d_0$ are strictly positive; it is easily checked that matrices of the form $\bd I+\bd u\bd v\tr$ are singular if and only if $\bd u\tr\bd v = -1$  %diverso => non singolare
which, in our context, is impossible because the elements of both $\bd a$ and $\bd d_0$ are strictly positive.
Using the fact that $(\bd I+\bd u\bd v\tr)^{-1}$ = $\bd I-\bd u\bd v\tr/(1+\bd u\tr\bd v)$
and some additional algebraic computations (see the Appendix for details), we have that
$$
\bd D_{\bdl\phi} = \diag(\bd d_0)^{-1}\left[\bd I - \bd a(\diag(\bd d_0)^{-1}\bd\phi+\diag(\bd d_1)^{-1}\bd\tau)\tr/g \right]
\diag(\bd d_1),
$$
with $g = 1+\bd a\tr\diag(\bd d_0)^{-1}\bd\phi$.

\begin{figure}[htb!]
\centering
 \includegraphics[width=0.9\textwidth]{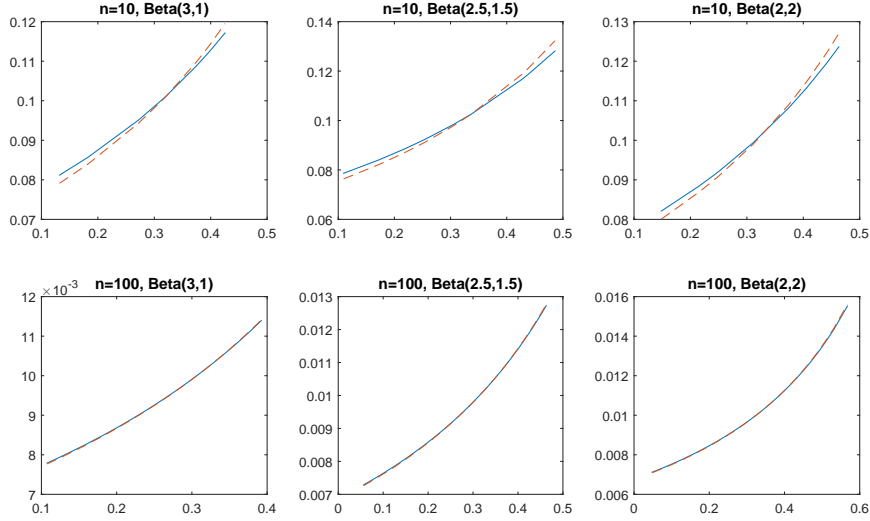}
\caption{\label{Fig1} \it Solid lines: plots of $\tau_i$ (vertical axis) as a function of $\phi_i$ (horizontal axis); dotted lines: plots of $\tau_i^{(c)}$ with respect to $\phi_i$.}
\end{figure}

When $\bd n\neq \bd 1$, as in \cite{Liu2017} (paragraph before their equation (6)), to account for ties, define $\omega_i$ = $\tau_i/n_i$; then, an implicit equation similar to (\ref{Inplicit}) with $\bd\tau$ replaced by the vector with elements $\omega_i$, $i=1,\dots,s$, is easily derived: start from \eqref{Inplicit}, replace $\bd 1$ with $\bd n$ and divide the $i$th equation by $n_i$.
\begin{remark}
Proposition \ref{IFT} indicates that $\bd\tau$ is uniquely determined by $\bd\phi$ and thus $\bd\beta$; though no explicit equation exists, once an estimate of $\bd\beta$ is available, an estimate of $\bd\tau$ is uniquely determined from (\ref{Inplicit}).
In addition, equation (\ref{Hyper}) implies that, element-wise, the relation between $\bd\phi$ and $\bd\tau$ is strictly increasing and has a simple functional form. It may be interesting to note that, because the conditional MLE of the population size in the $i$th stratum may be computed as $N\tau_i^{(c)}$ where $\tau_i^{(c)}$, the relative size of the $i$th stratum in the case of only one individual captured per stratum, takes the form
$$
\tau_i^{(c)}= \frac{1}{1-\phi_i}\left(\sum_{i=1}^s\frac{1}{1-\phi_i}\right)^{-1},
$$
which is again the increasing branch of an hyperbola.
Values of $\tau_i$ and $\tau_i^{(c)}$ as functions of $\phi_i$ are displayed in Figure \ref{Fig1} by sampling the $\bd\phi$ from three different Beta distributions for $n=10$ and $n=100$. It emerges that, when $n$ is sufficiently large, the two methods give almost equivalent results. When $n$ is small, the conditional MLE assigns a lower proportion of never captured individuals to strata with smaller $\phi_i$s relative to the approach presented here.
\end{remark}
\subsection{Updating $N$} \label{sec:Implicit}
This may be performed by a one-dimensional Fisher-scoring step based on the derivatives of the log-gamma function $\log[\Gamma(v)]$ to compute the score and $_oF_N$, the observed information for $N$.
These quantities may be expressed as
\begin{eqnarray*}
s_N &=& D_{lg}(1,N+1) - D_{lg}(1,N-n+1) + \log(\phi),\\
_oF_N &=& D_{lg}(2,N-n+1)-D_{lg}(2,N+1),
\end{eqnarray*}
where $D_{lg}(k,v)$ is the $k$th derivative of $\log[\Gamma(v)]$.
An updating rule may have the form
$$
N^{(u+1)} = N^{(u)} + a_u s_N/F_N,
$$
where $a_u$ denotes the step length, for instance one may set $a_1$ = 0.5 and let $a_u$ tend to 1 as $u$ increases.
\section{Asymptotic results}\label{sec:asymp}
Proposition \ref{IFT} implies that one can replace the original likelihood with its profile with respect to $N$ and $\bd\beta$ by replacing $\bd\tau$ with $\hat{\bd\tau}(N,\bd\beta)$, its MLE for a given value of the arguments.
Recalling that the parameter vector $\bd\psi$ = $(N, \bd\beta\tr)\tr$ does not contain the nuisance parameters $\bd\tau$, to derive the asymptotic distribution of $\hat{\bd\psi}$ we can use the results in \cite{Liu2017}, Supplementary Material, with minor adaptations which are described in the Appendix.
It follows that, as $N_0$, the true value of the population size, goes to infinity, the MLE is consistent and
$$
\sqrt{N_0} (\hat{\bd\psi}-\bd\psi) \overset{d}{\to} N(\bd 0, \bd F^{-1}),
$$
where $\bd F$ is the asymptotic information of the profile likelihood, as explained below.
\subsection{The information matrix}
The information matrix used in the Fisher scoring algorithm, in addition to being $N$ times larger than $\bd F$, assumes that $\bd\tau$ is known; to take into account uncertainty about $\bd\tau$, we need to compute the information matrix of the profile likelihood (see for instance \cite{PaceSalvan}, Sec. 4.6). The following argument may be used to compute an explicit expression for the expected information matrix: first, use the results in  \cite{Liu2017} who derive an approximation for $F_{NN}$ and show that $\bd F_{N\bdl\beta}$ converges to a vector of 0s when $N_0$, the true population size, tends to infinity.
In addition, because Lemma \ref{FInf} in the Appendix implies that the expected value of all components of the observed information involving derivatives of $\bd\Phi,\: \bd D_i$, $i=1,\dots ,s$, with respect to the elements of $\bd\beta$ are 0, when differentiating $\bd s_{\bdl\beta}[\bd\beta,\hat{\bd\tau}(\bd\beta)]$ with respect to $\bd\beta\tr$, we can restrict attention to terms containing $\bd p_i,\:\bd\phi,\:\bd\tau$.
Direct calculations show that
\begin{align*}
F_{NN} &= \frac{1-\phi}{\phi},\\
\bd F_{\bdl\beta\bdl\beta} &= \sum_{i=1}^s \tau_i\bd D_i\tr\diag(\bd p_i)^{-1}\bd D_i
+\bd\Phi\tr\left[\left(\bd\tau \bd\phi\tr /\phi - \bd I\right) \bd D_{\bdl\phi}
+\bd\tau\bd\tau\tr/\phi \right]\bd\Phi.
\end{align*}
\begin{remark}
Though the latter expression may appear hard to compute when $s$ is very large, the results in Proposition \ref{IFT} provide an explicit expression for $\bd D_{\bdl\phi}$ which, in addition, has a structure that makes matrix multiplications simpler.
Though an explicit expression for the observed information matrix of the profile likelihood is very hard to compute, an accurate numerical approximation may be obtained by computing the numerical derivative of  $\bd s_{\bdl\beta}[\bd\beta,\hat{\bd\tau}(\bd\beta)]$ with respect to $\bd\beta\tr$.
\end{remark}
\subsection{Confidence intervals for $N$}
The previous results could be used to construct a confidence interval for $N$; however, this would lead to an interval which is symmetric around $\hat N$, the MLE, while it is likely that uncertainty is larger on the right hand side.
{Confidence intervals} for $N$ based on the conditional likelihood were proposed by \cite{Bart2006} among others; an approach for constructing confidence intervals based on \cite{Owen88}'s notion of Empirical Likelihood is proposed by \cite{Liu2017}.

Let $\hat N$ and $\hat{\bd\beta}$ denote the unconditional MLEs of the corresponding parameters; let also $\hat{\bd\beta}_N$ denote the corresponding estimate when the population total is kept fixed at $N$; define
$$
D_N = 2\left[\tilde L(\hat N,\:\hat{\bd\beta}) - \tilde L(N,\:\hat{\bd\beta}_N)\right];
$$
then the following results holds.
\begin{proposition}
The asymptotic distribution of $D_N$ converges to $\chi_1^2$ as $N_0$ grows to infinity.
\end{proposition}
\noindent{\sc Proof.} The proof is similar to that of Theorem 1 in \cite{Bart2006} which uses, essentially, two steps: (i) show that $D_N$ may be approximated by a constrained quadratic function of the parameter vector $\bd\psi$ and (ii) exploiting the asymptotic normal distribution of the ML estimates.
\section{Applications}\label{sec:application}
\subsection{Bacterial meningitis}
The data were produced by the research unit on Infectious Diseases of Lazio, a region which includes the metropolitan area of Rome. Cases were recorded by four different sources:
\begin{itemize}
\item{\em HSS}: hospital surveillance of bacterial meningitis;
\item{\em NDS}: the mandatory infectious diseases notifications;
\item{\em LIS}: the laboratory information system;
\item{\em HIS}:
hospital information system.
\end{itemize}

Records appearing in the four lists from 2001 to 2005 were combined into a single archive. For a detailed description of the context, see \cite{PGRossi2009}. For each of the $n$ = 944 patients appearing in at least one list, several covariates were recorded. As in \cite{bartolucci2018latent}, we restrict our attention to the following covariates that were considered to be the most relevant:
\begin{itemize}
\item{\em Age}: binary variable equal to 1 for up to 1 year old and 0 otherwise, because the incidence of meningitis is much higher among very younger children;
\item{\em Aez}: binary variable for the recorded type of bacteria, which is equal to 1 for Pneumococcus, Meningococcus or Tuberculosis and 0 otherwise;
\item{\em Year}: year of first appearance in a list; this is included because the way of functioning of certain lists has evolved during the study period.
\end{itemize}

Here we fit the log-linear analog of the best marginal model according to \cite{bartolucci2018latent}, who however used a conditional likelihood approach.
There are two latent classes whose weights are assumed to depend on Age and Aez, the conditional logits depend linearly on Year and, in addition, there are log-linear interactions between lists (1,2) and (2,4) not depending on the latent.
To compare the estimated capture probabilities by strata of the marginal model relative to the log-linear one, the Kulback-Leibler divergence times the population size (as estimated by the log-linear model) was used. Considering that the maximum divergence is about 1.7 and that the correlation coefficient between the probability of being never capture by strata in the two approaches is about 0.9998, we may conclude that the two models produce almost the same estimates. However, some differences emerge in the estimate of the population size which is 1,359 according to the log-linear model against 1,386 in the marginal model. The 95\% confidence interval ranges from 1127 to 1828, against 1142 to 1875 in the marginal model. These differences are more likely to be due to the fact that the new approach takes into account the marginal distribution of the covariates.
\subsection{Deer mouse data}
The data, included in the {\tt R} package {\tt VGAM}, are about a small rodent native to North America, called deer mouse.
Overall, 38 individuals were trapped at least once during 6 consecutive nights. Available covariates are sex, age (binary), and weight in grams.
According to \cite{otis1978statistical}, page 32, who analyzed the data, model $M_b$, assuming that already captured animals are more likely to be captured again, fits well, giving $\hat N$ = 41.
Based on the assumption of normality, they estimated a confidence interval between 35 to 47, where clearly the lower limit has to be corrected to 38.

Model search was limited by the fact that $n$, the number of subjects captured at least once, is small relative to the number of occasions (lists). Thus in all models we assume the existence of two latent classes and that the capture probability conditionally on the latent, covariates,
and capture history is constant over time. Several models with latent weights or conditional distributions depending on covariates were considered; among the  covariates, only sex seems to have a substantial effect on the latent distribution. This model has just four parameters which determine the probability of being in latent class 2 for males, how this probability changes for females and the capture probability for subjects in latent class 1 and 2. Capture probability for latent class 2, which includes mostly males, is about 0.83 against 0.36 for latent class 1. The likelihood ratio for testing the dependence of latent weights on sex is 7.8 with 1 d.o.f., thus highly significant.

The latent class version of the $M_b$ model has two additional parameters which are highly significant; however, two parameter estimates are near the boundary and the corresponding standard errors are very large indicating that the asymptotic approximations may not be reliable. A constrained version of the same model where the difference, on the logit scale, between the probability of capture in the first occasion and that of recapture is constant across latent classes, has more reasonable estimates, except that, for females, the probability of being in latent class two is very close to 0. Both latent types seem to be trap lovers: the estimates of the capture probabilities in the first occasion are 0.26 and 0.74 respectively and they jump to 0.45 and 0.86 when already captured in a previous occasion. Taking this as our final model, the point estimate of $N$ equals 42 and the 95\% confidence interval goes from 38 to 60; the fact that the upper limit is so much larger than the one in \cite{otis1978statistical} cannot be attributed to the refinements in the model and are plausibly due to the fact that the uncertainty in the size of $N$ is much larger on the upper side, a feature which cannot be captured by the normal approximation.
\section{Concluding remarks}
A natural question is what do we gain by modeling the marginal distribution of the covariates. The main role of the $\bd\tau$ parameter is to allow us defining the average probability that a subject with unspecified covariate values is never captured. More importantly, it allows us to deal with the overall population size $N$ directly, rather than to estimate the population size within each strata, say $N_i$, and then define $N=\sum_{i=1}^s N_i$.
Moreover, the simulation results in \cite{Liu2017} indicate that the present approach, relative to the one where we condition to the observed covariates and to being captured at least once, leads to more efficient estimates of the population size and to shorter width of the corresponding confidence intervals. A limited simulation that we performed indicates that the same holds within our extended context.

As far as we know, the idea of modeling the marginal distribution of the covariates appeared in \cite{Liu2017} for the first time; the previous literature dealing with a collection of populations, see for instance \cite{StangPvdH} and \cite{Hggins1989}, use as parameters the size of individual populations. Though \cite{Sanathanan}, Section 5, shows that the unconditional ML estimate are consistent and asymptotically normal if the relative size of each stratum does not tend to 0 as $\sum_{i=1}^s N_i$ tends to infinity, unless $N$ is sufficiently large, the estimates within strata may be problematic as explained below.

Unless there is a certain amount of under-count in each stratum, that is  $\hat N_i>n_i$, estimates of $\bd\beta$ may be close to the boundary.
When the marginal weights of the strata are not taken into account,  each $N_i$ may be estimated separately and Lemma 1 in \cite{Sanathanan} implies that, for the condition $\hat N_i>n_i$ to be satisfied, we must have $\hat \phi_i> 1/(n_i+1)$. It can be easily seen that, unless $n_i$ is sufficiently large, $\hat N_i = n_i$ because the score with respect to $N_i$ is negative at $N_i=n_i$. For instance, when $n_i=1$, from the properties of the $D_{lg}$ function, it can be shown that the score at $N_i=n_i$ equals $1+\log(\phi_i)$ which is negative unless $\phi_i > 0.3679$; when $n_i=2$, the score equals $1.5+\log(\phi_i)$ which, again, is negative unless $\phi_i>0.2231$. In practice, unless the under-count is sufficiently large relative to $n_i$, it cannot be detected during model estimation.
\section*{Acknowledgments}
The authors would like to thank Yukun Liu and Yang Liu, from the East China Normal University, for helpful discussions about the \cite{Liu2017} paper. Yosef Rinott, from the Hebrew University of Jerusalem, helped on certain mathematical issues.
\section*{Appendix}
\subsection*{The relation between $\bd\tau$ and $\bd\phi$}
An updating equation for $\bd\tau$ may be obtained as follows; (i) differentiate $L(\bd\tau)$ with respect to $\bd\tau$ gives $(N-n)\bd\phi/\phi +\diag(\bd\tau^{-1})\bd n-\gamma\bd 1$ = $\bd 0$;  (ii) left multiply by $\bd\tau\tr$ and solving gives $\gamma$ = $N$; (iii) substitute for $\gamma$ and left-multiply by $\diag(\bd\tau)$ gives equation (\ref{tau-it}).

From now on we restrict attention to the case of $\bd n$ = $\bd 1$; in the general case, the relations derived below hold between $\phi_i$ and $\omega_i$ = $\tau_i/n_i$.
The element-wise version of (\ref{tau-it}) at convergence implies that $\tau_iN\phi$ = $\phi+(N-n)\phi_i\tau_i$;
solving for $\tau_i$ gives (\ref{Hyper}).

To obtain an expression for $\bd D_{\bdl\phi}$, we first differentiate (\ref{Inplicit}) with respect to $\bd\phi\tr$ by using the following partial results
\begin{align*}
& \frac{\partial\phi}{\partial\bd\phi\tr} = \bd\phi\tr\frac{\partial\bd\tau}{\partial\bd\phi\tr}+\bd\tau\tr,\quad
\frac{\partial\phi\bd\tau}{\partial\bd\phi\tr} = \bd\tau\left(\bd\phi\tr\frac{\partial\bd\tau}{\partial\bd\phi\tr}+ \bd\tau\tr\right)+\phi \frac{\partial\bd\tau}{\partial\bd\phi\tr},\\
& \frac{\partial\:\diag(\bd\phi)\bd\tau}{\partial\bd\phi\tr} = \diag(\bd\phi) \frac{\partial\bd\tau}{\partial\bd\phi\tr}
+\diag(\bd\tau).
\end{align*}
With these tools, the derivative of the implicit equation above is
$$
N\bd\tau\left(\bd\phi\tr\bd D_{\bdl\phi}+\bd\tau\tr\right)+N\phi\bd D_{\bdl\phi} -\bd 1\left(\bd\phi\tr\bd D_{\bdl\phi}+\bd\tau\tr\right)- (N-n)\left(\diag(\bd\phi)\bd D_{\bdl\phi}+\diag(\bd\tau)\right)=\bd 0;
$$
now collect together all the terms that left multiply $\bd D_{\bdl\phi}$ and move everything else to the right hand side. Equation (\ref{eq:Dphi}) follows by simple algebra.

Let $\bd d$ = $\bd 1 - (N-n)\bd\phi/(N\phi)$ and $\bd t$ = $\diag(\bd d)^{-1}(\bd\tau-\bd 1/N)/\phi$; the following lemma holds.
\begin{lemma}\label{diti}
If the elements of $\bd\phi$ and $\bd\tau$ are strictly positive, then  the elements of $\bd d$ are positive and those of $\bd t$ are non-negative provided that $N\geq n$.
\end{lemma}
{\sc Proof}: Left multiply equation (\ref{Inplicit}) by $\diag(\bd\tau)^{-1}/(N\phi)$, rearrange terms and let $\dot{\bd\tau}$ denote the vector with elements $1/\tau_i$; we have that
$$
\bd 1 - (N-n)\bd\phi/(N\phi) = \dot{\bd\tau}/N >\bd 0,
$$
which implies that the elements of $\bd d$ are strictly positive. Equation (\ref{Inplicit}) also implies that $\bd\tau-\bd 1/N$ =  $(N-n)\diag(\bd\tau)\bd\phi /(N\phi)$; because the elements of the latter expression are non-negative, $\tau_i\geq 1/N$ for all $i$, which also implies that the elements of $\bd t$ are non-negative.
\subsection*{Asymptotic approximations}
\cite{Liu2017}'s approximations hold also in our context if the assumed latent class model is identifiable in a neighborhood of the true $\bd\beta$. Though no exact result is available for identifiability of latent class models where marginal weights and conditional distributions depend on covariates, identifiability might be tested numerically as described by \cite{Forcina08}: for a sample of $\bd\beta$ points drawn in a neighborhood of the MLE, one may check that the derivative of the canonical parameter of an arbitrary $\bd p_i$ with respect to $\bd\beta\tr$ is away from being singular.
A simpler alternative is to check that the expected information matrix $\bd F_{\bdl\beta\bdl\beta}$ is positive definite.
\subsection*{Score vector and expected information matrix}
Recall that, if $\bd f(\bd x)$ is a vector of functions such that $\bd f\tr\bd 1=1$, the following identities hold
$$
(a) \:\frac{\partial\bd f\tr}{\partial\bd x}\bd 1 =\bd 0, \quad (b)\:
\frac{\partial^2 \bd f\tr}{\partial x_h\partial x_j}\bd 1=\bd 0.
$$
Consider the score vector $\bd s_{\bdl\beta}$ in a capture-recapture context and let $\tilde{\bd D}_i$ denote the matrix $(\bd d_i\:\: \bd D_i\tr)\tr$, where $\bd d_i$ denotes the $i$th column of $\bd\Phi\tr$; recall that E$(N-n)=N\phi$ and E$(\bd y_i)$ = $N\tau_i\bd p_i$. Define $\bd a_i$ = $(\phi\:\: \bd p_i\tr)\tr$, $\bd b_i$ = $(\tau_i(N-n) \: \: \bd y_i\tr)\tr$, $\bd u_i$ = $\diag(\bd a_i)^{-1}\bd b_i$ and recall that $\tilde{\bd p}_i$ denotes the vector of capture probabilities that includes the missing cell; note that E$(\bd u_i)$ is proportional to the unitary vector.
\begin{lemma} \label{FInf}
$(i)$  E$(\bd s_{\bdl\beta})$ = $\bd 0$ and
$(ii)$ E$[\partial^2 \tilde{\bd p}_i\tr / (\partial\beta_h\partial\beta_j)\bd u_i ] = 0$.
\end{lemma}
{\sc Proof}
The expression of the score may be written as
$$
\bd s_{\bdl\beta} = \frac{N-n}{\phi}\bd\Phi\tr \bd\tau + \sum_{i=1}^s\bd D_i\tr
\diag(\bd p_i)^{-1}\bd y_i = \sum_{i=1}^s\tilde{\bd D}_i\tr \bd u_i,
$$
then $(i)$ and $(ii)$ follow respectively from $(a)$ and $(b)$ above by noting that the $j$th column of $\tilde{\bd D}_i$ is equal to the derivative of $\tilde{\bd p}_i$ with respect to $\beta_j$.

The practical implication of $(ii)$ is that, when computing the expected information matrix, terms obtained by differentiating $\tilde{\bd D}_i$ may be neglected.

\bibliographystyle{apalike}
%\nocite{*}
\bibliography{CaRi2}

\end{document}